\documentclass[lettersize,journal]{IEEEtran}

\usepackage{tikz}
\usepackage{amsmath,amsfonts}
\usepackage{algorithmic}
\usepackage{array}
\usepackage[caption=false,font=normalsize,labelfont=sf,textfont=sf]{subfig}
\usepackage{textcomp}
\usepackage{stfloats}
\usepackage{url}
\usepackage{verbatim}
\usepackage{graphicx}
\usepackage[style=ieee]{biblatex}
\usepackage{glossaries} 
\usepackage{tabularx}
\usepackage[T1]{fontenc}

\def\BibTeX{{\rm B\kern-.05em{\sc i\kern-.025em b}\kern-.08em
    T\kern-.1667em\lower.7ex\hbox{E}\kern-.125emX}}
\usepackage{balance}

\usepackage{amssymb} 

\newacronym{acr_csp}{CSP}{Closest String Problem}
\newacronym{acr_qpu}{QPU}{Quantum Processing Unit}
\newacronym{acr_qpus}{QPUs}{Quantum Processing Units}
\newacronym{acr_qubo}{QUBO}{Quadratic Unconstrained Binary Optimization}
\newacronym{acr_sdk}{SDK}{Software Development Kit}

\bibliography{references/ref_toc}
\bibliography{references/ref_qc}
\bibliography{references/ref_inft}
\bibliography{references/ref_qm}
\bibliography{references/ref_opt}
\bibliography{references/ref_general}

\newcommand\copyrighttext{%
  \footnotesize \textcopyright This work has been submitted to the IEEE for possible publication. Copyright may be transferred without notice, after which this version may no longer be accessible.}
\newcommand\copyrightnotice{%
\begin{tikzpicture}[remember picture,overlay]
\node[anchor=south,yshift=10pt] at (current page.south) {\fbox{\parbox{\dimexpr\textwidth-\fboxsep-\fboxrule\relax}{\copyrighttext}}};
\end{tikzpicture}%
}

\begin{document}

\title{Quantum Annealing Solutions for the \acrlong{acr_csp} with D-Wave Systems}
\author{Chandeepa Dissanayake
\thanks{The author is with the Department of Statistics \& Computer Science, University of Kelaniya, Dalugama 11300, Sri Lanka (email: chandeepadissanayake@gmail.com)}}

\markboth{IEEE Transactions on Computers}%
{Quantum Annealing Solutions for the \acrlong{acr_csp} with D-Wave Systems}

\maketitle
\copyrightnotice

\begin{abstract}
The \acrlong{acr_csp} is an NP-complete problem which appears more commonly in bioinformatics and coding theory. Less surprisingly, classical approaches have been pursued with two prominent algorithms being the genetic algorithm and simulated annealing. Latest improvements to quantum computing devices with a specialization in optimization tasks such as D-Wave systems, suggest that an attempt to embed the problem in a model accepted by such systems is worthwhile. In this work, two \acrshort{acr_qubo} formulations have been proposed, with one being a slight modification over the other. Subsequently, an evaluation based on a few simple test cases had been carried out on both formulations. In this regard, the D-Wave annealers have been used, while providing guidelines for optimality on certain platform-specific concerns. For evaluation purposes, a metric termed Occurrence Ratio (OR) has been defined. With minimal hyperparameter tuning, the expected solutions were obtained for every test case and the optimality was guaranteed. To address practical and implementation issues, an inherent decomposition strategy based on the possibility of having substrings has been elucidated to accommodate the restricted qubit count. Conclusively, the need for further investigation on tuning the hyperparameters is emphasized.
\end{abstract}


\section{Introduction}
\IEEEPARstart{F}{ormal} languages are an important concept in automata theory. Informally, such a formal language is defined as a set of strings constructed using the characters of a finite alphabet\cite{sipser_introduction_2006}. In this context, different scenarios emerge in which the objective is to determine a representative string that exhibits the highest degree of similarity to a given set of strings. The concept of similarity and the need to quantitatively measure it, gives rise to different objective functions and thus different distance measures for determining the difference between two strings. Note that different distance measures will yield different representative strings.\par

For any two strings of equal length, the Hamming distance is one of the most fundamental distance measures, which has its roots in coding theory\cite{hamming_error_1950}. Hamming distance is the number of positions at which the corresponding symbols are different for any two strings of equal length\cite{waggener_pulse_1995}. More formally, it can be defined with the Kronecker delta $\delta_{ab}$ as follows. Given two strings $s_{1}, s_{2}$ of length $m$ over the alphabet $\Sigma$, the Hamming distance $d(s_{1}, s_{2})$ is given by, 
\begin{equation} \label{eq:intro:hamming_distance}
    d(s_{1}, s_{2}) = \sum_{i=1}^{m}f(s_{1i},s_{2i})
\end{equation}
where $s_{xi}$ denotes the $i^{th}$ character of the string $s_{x}$ ($x = 1$ or $x = 2$ in this definition), and the function $f$ is defined as,
\begin{equation} \label{eq:intro:hamming_distance_f}
    f(c_{1}, c_{2}) = 1 - \delta_{c_{1}c_{2}} = 
        \begin{cases}
            0, & \text{if } c_{1} = c_{2}\\
            1, & \text{if } c_{1} \neq c_{2}
        \end{cases}
\end{equation}
Note that $c_{1}, c_{2}$ are two variables. In Equation (\ref{eq:intro:hamming_distance}), we set $c_{1} = s_{1i}$ and $c_{2} = s_{2i}$ for each $i$.\par

The problem of determining the representative string with the aid of Hamming distance as the distance measure is known as the \acrfull{acr_csp}. Formally, elaborating on the definitions in \cite{lanctot_distinguishing_2003, li_closest_2000}: given a set of strings $S = \{s_{1}, s_{2}, s_{3}, ..., s_{n}\}$ where each string $s_{x}$ is defined over the alphabet $\Sigma$ and is of length $m$, the goal is to determine a string $s_{M}$ which minimizes $k$ such that for each string $s_{x} \in S$,
\begin{equation} \label{eq:intro:csp_condition}
    d(s_{x}, s_{M}) \le k
\end{equation}\par

\acrshort{acr_csp} is proven to be NP-hard\cite{lanctot_distinguishing_2003}. A special case of the decision problem of \acrshort{acr_csp} called the hitting string problem has been proven to be NP-complete\cite{fagin_generalized_1974}. Therefore, the decision problem of \acrshort{acr_csp} is NP-complete. Thus, exploring algorithms with a lower time complexity for \acrshort{acr_csp} does not purely rely in the interests of applications of the problem. From a theorist’s point of view, such investigations could lead to better insights about computationally hard problems and possibly allow further intuitions about unsolved problems such as P vs. NP question.\par

Coding theory is one of the major fields in which \acrshort{acr_csp} has its applications, mostly in error correction\cite{gasieniec_efficient_1999, frances_covering_1997}. A reader familiar with the bioinformatics might notice that closest string has to be determined in designing genetic probes and drug target identification \cite{lanctot_distinguishing_2003}. Algorithm presented in \cite{stormo_identifying_1989} for the identification of protein binding sites presents a direct application of finding the closest string representation. Additionally, interesting applications might be found for different problems in the context of Finite and Push-Down Automata in automata theory. \cite{han_closest_2021} discusses a generalization of \acrshort{acr_csp}, called the Closest Substring Problem in the context of regular languages.\par

Classically, this is one of the problems that had been studied extensively. One of the earliest approximation algorithms, called the Largest Distance Decreasing algorithm has been presented in \cite{liu_largest_2004}, and determines solutions in polynomial time. The greedy heuristic algorithm in \cite{vilca_recursive_2022} chooses the character based on the global evaluation of the Hamming distance and the choice at a given index affects the choice of the symbol for the next. Wave function collapse techniques have been studied in \cite{noauthor_heuristic_nodate} as another heuristic based approach, using the idea of entropy in the original WaveFunctionCollapse constraint \cite{gumin_wave_2016}. Additionally, a hybrid metaheuristic is discussed in \cite{mousavi_hybrid_2011}. For guaranteeing the optimality, a recursive exact method has been presented and has been compared against an integer programming formulation of the CSP\cite{vilca_recursive_2022}. Whenever the alphabet is of size 2, \cite{liu_exact_2011} presents an exact algorithm called Distance First Algorithm. For the other cases, a polynomial heuristic has been introduced and it is claimed that there is a possibility of obtaining a nearly optimal solution in a reasonable time. Furthermore, \cite{yuasa_designing_2019} presents a set of implementations of the existing algorithms and introduces a parameterized algorithm for the binary case of CSP.\par

Evidently, none of these approaches were successful in providing an efficient algorithm for the \acrshort{acr_csp} in general. This could very well be due to the intrinsic difficulty in the problem itself, or due to the inherent limitations in the model of computation we rely upon. For example, despite the integer factorization problem not being proven to be NP-complete and exponential speedup hasn’t been obtained, Shor’s algorithm built on top of a unitary circuit model of quantum computation demonstrated superpolynomial advantage over every existing classical algorithm\cite{shor_algorithms_1994}. Accordingly, it is worthwhile to investigate alternative models of computation for problems that are either NP-complete or appear to be hard.\par

Adiabatic quantum computation is a quantum computational model that has seen interest in the recent years, particularly due to the development of devices that could leverage the process of quantum annealing for solving different problems, and especially optimization problems. Quantum annealing is involved in the minimization of an energy function through adiabatic evolution, and it requires the re-formulation of the existing problem definition as a Hamiltonian that defines the energy of a quantum system. The reformulation of the problem can be modeled either as an Ising problem or as a \acrfull{acr_qubo} problem, with the latter simply being a transformation of the former. Retrospectively, a variety of NP-complete problems have been formulated in this way\cite{lucas_ising_2014} and had been attempted using different annealing devices. In conclusion, it is worthwhile to investigate such a formulation for the \acrshort{acr_csp}.\par

The prime outcome of this work is two \acrshort{acr_qubo} formulations for the \acrshort{acr_csp}. The formulated \acrshort{acr_qubo} models have been evaluated on a limited number of simple test cases, where the D-Wave systems were used. D-Wave Systems is a vendor of quantum annealing devices and recently, they have allowed public access to their \acrfull{acr_qpus} on the cloud supplemented by an SDK. Subsequently the need for utilizing the number of available qubits and the techniques of achieving it will be discussed. Finally, the need for hyperparameter tuning has also been emphasized.\par

\section{Preliminaries}
    \subsection{Ising Model}
        An Ising model is an abstract mathematical model which usually has a large, but finite number of states. It has been used to describe different physical systems and the properties of them such as ferromagnetism in statistical mechanics. In fact, it is convenient to apprehend an Ising model as a lattice structure\footnote{Referring to how the ions are arranged in the crystal structure of a metal that exhibits magnetism (Iron in this case)} in which there are lattice sites where the unit cells of the lattice can be located in either spin up or spin down state. In this context, it is important to view a lattice as a graph rearranged in 3-dimensions. Formally, the Ising model is defined as follows.\par

        Consider a lattice structure (as described above) where the set of lattice sites are given by $\wedge$. Every lattice site $k \in \wedge$ has a set of adjacent lattice sites. For each lattice site $k \in \wedge$, there exists a discrete variable $\sigma_k$ such that $\sigma_k \in \{-1, 1\}$, representing the spin. The spin configuration $\sigma = \{\sigma_k\}_{k \in \wedge}$ is an assignment of spin value to each site. For any 2 adjacent sites, there is an interaction $J_{ij}$ where $i,j \in \wedge$. There is an external parameter denoted by $h$, which is typically an external magnetic field that interacts with the lattice site. Accordingly, for a given configuration, the energy is given by,
        \begin{equation}\label{eq:prelim:ising_1}
            H(\sigma) = -\sum_{<ij>}J_{ij}\alpha_{i}\alpha_{j} - h \sum_{j}\sigma_{j} 
        \end{equation}
        The $(-)$ sign of the second term of Equation (\ref{eq:prelim:ising_1}) is conventional. Furthermore, note that there are simplifications of this formulation, depending on the specific problems that are being modelled. For a detailed derivation and and explanation refer to \cite{ising_beitrag_1925}.\par

        \subsubsection{Transverse Field Ising Model}
            The quantum mechanical description of the above Ising model is called the Transverse Field Ising model. In here, the interactions $J_{ij}$ between lattice sites $i$ and $j$ are determined by the spin projections of the involved lattice sites spins along the $z$-axis. These interactions are also affected by the external magnetic field (which accounted for the external parameter $h_j$ in the classical model), which acts perpendicular to the $z$-axis in this case, say along $x$-axis. This necessity of the perpendicularity in this setup, renders the spin projection along $z$-axis and the $x$-axis to be non-commuting observables. Consequently, the classical model cannot explain this setup, thus requiring the replacement of the spins with Pauli matrices associated with spin-$1/2$ observables. Accordingly, the energy of any spin configuration is given by the quantum Hamiltonian,
            \begin{equation}\label{eq:prelim:ising_2}
                H^{'} = -J\left\{\left(\sum_{<i, j>}Z_{i}Z_{j}\right) + g\sum_{j}X_j \right\}
            \end{equation}
            where, $Z_{i}$ and $X_i$ are Pauli matrices (as described already), and $J$ is simply a prefactor, whereas $g$ represents a coefficient which determines the relative strength of the external magnetic field applied when compared to the interactions between neighbouring sites.\par

            In one of the foundational papers, quantum annealing has been proposed by introducing quantum fluctuations to the simulated annealing paradigm. It was tested by using the transverse field Ising model\cite{kadowaki_quantum_1998}.
    
    \subsection{\acrfull{acr_qubo}}
        \acrlong{acr_qubo} problem is an NP-hard combinatorial optimization problem which attempts to determine a minimum value for a function defined on a binary vector space, along with the corresponding binary vector which results in the minimum value. Formally, given an upper triangular matrix $Q^{n \times n}$, the objective is to determine a binary vector $x^{*} \in \{0, 1\}^{n}$ such that $argmin\, f(x) = x^{*}$ where, 
        \begin{equation} \label{eq:prelim:qubo_1}
            f(x) = x^{T}Qx = \sum_{i = 1}^{n}\sum_{j = 1}^{n} Q_{ij}x_{i}x_{j}
        \end{equation}
        for any $x \in \{0,1\}^{n}$.\par

        The \acrshort{acr_qubo} problem displays a close resemblance with the Ising model formulation. In fact, any such Ising model formulated in terms of spins $s_{\alpha}$ can be transformed to a \acrshort{acr_qubo} problem, replacing each spin by a binary variable, by applying the following transformation.
        \begin{equation} \label{eq:prelim:qubo_2}
            x_{\alpha} = \frac{s_{\alpha} + 1}{2}
        \end{equation}\par
        
        Generally, for most of the problems it is more convenient to develop the model as a \acrshort{acr_qubo} problem rather than an Ising formulation\cite{lucas_ising_2014}. Furthermore, by introducing coefficients/penalties for each an every component in the formulation, the penalty Hamiltonian is obtained and it is used to embed the problem in the annealing devices. This is because such quantum annealing devices are capable of solving only unconstrained problems, and if there are constraints embedded into the Hamiltonian, the only possibility is to drastically increase the energy on each constraint violation\cite{lucas_ising_2014}.
    
    \subsection{D-Wave Quantum Annealers}
        D-Wave is a vendor which allows public access to a set of quantum annealers through a cloud called Leap. They provide a \acrfull{acr_sdk} called Ocean \acrshort{acr_sdk}, along with a variety of other toolkits which suits different purposes from scientific research to commercial grade applications. In this work, D-Wave Leap cloud was utilized through the use of their Ocean \acrshort{acr_sdk} to test and evaluate the \acrshort{acr_qubo} formulations.\par

        In order to solve an optimization problem using the annealers in D-Wave systems, it is required to embed the problem in to the \acrshort{acr_qpus} according to their topology as the first step. Once the problem is formulated as an Ising model or a \acrshort{acr_qubo} problem, it is required to map it into the \acrshort{acr_qpu}. In the very early D-Wave \acrshort{acr_qpus}, this mapping was achieved through the Chimera graph\cite{johnson_scalable_2010}. The process of embedding problem variables into the \acrshort{acr_qpu}, which is called minor-embedding, was to be manually accomplished. With the recent upgrades up to the Advantage QPUs, the primary topology was also upgraded to Pegasus graphs which has a slightly different internal arrangement of couplers\cite{boothby_next-generation_2020}. Furthermore, the D-Wave Ocean \acrshort{acr_sdk} was supplemented with routines to implicitly perform minor embedding, thus eliminating the need for the user to be aware of the internal architecture of the \acrshort{acr_qpu} being accessed. An interested reader may refer to \cite{noauthor_getting_nodate} for further explanation.\par

        Both of the above topologies are incapable of mapping any given problem formulation as a one-to-one mapping from problem variables to physical qubits on the \acrshort{acr_qpu}. In such scenarios, several physical qubits are chained together to represent a single problem variable during the minor-embedding stage. Whenever the problem is solved in the QPU, these chained qubits are constrained to have the same value in every solution. In case this constraint is violated, the chain is said to be broken and the embedding does not represent the problem of interest anymore. During the annealing stage, the chains can be broken when attempting to minimize the energy function. ``chain\textunderscore strength" parameter is used to counter-act the tendency to break the chains\cite{systems_programming_2020}. A more detailed discussion and guidelines to set the value of this parameter can be found in \cite{systems_programming_2020}.\par

        The next step is to use a sampler to extract low energy states from the minor-embedded problem in the \acrshort{acr_qpu}. D-Wave provides different mechanisms for this based on the type of the device in which the problem is actually solved in, which they call the ``solver". Since the interest of this work is to attempt the problem on real \acrshort{acr_qpus}, ``QPU Solvers" have been utilized. For other types of solvers, refer to \cite{noauthor_workflow_nodate}. A sampler implements the process of sampling based on the solver that we have opted to use.\par

        \acrshort{acr_qpus} are probabilistic by design. D-Wave annealers are of no exception for this. Hence, the most sensible approach of obtaining a valid solution is to run the given problem as arbitrarily many times as possible and utilize the statistical trends observed, to determine the solution. A single run is called a ``read" or an ``annealing cycle". Multiple such reads increases the diversity of the solution space and in turn allows to determine the probability of obtaining the respective solution, which would not have been possible otherwise. In this context, when submitting any problem to the ``solver", D-Wave allows to specify a parameter called ``num\textunderscore reads" whose value denotes the number of annealing cycles \cite{systems_programming_2020}.\par

\section{Formulation of the \acrshort{acr_csp} as a \acrshort{acr_qubo} Problem}
    The problem is initially reformulated to a form which requires minimization of the sum of the Hamming distances between a candidate solution $s_{c}$ (a string over the alphabet $\Sigma$) and each string (over the alphabet $\Sigma$) in the given set $S$. This is transformed again into another form in which it is required to determine each symbol $s_{ci}$ of the candidate solution (the string $s_{c}$), such that the contribution to the sum of the Hamming distances by selecting the symbol $s_{ci}$ is the minimum among the sum of Hamming distances by selecting other symbols from a finite alphabet. Based on the use of Hamming distance as the distance measure, it is inferred that for each symbol in the candidate solution, the symbol which results in the minimum sum of Hamming distance lies within the subset $\Sigma_{i} \subset \Sigma$ comprised only with the symbols of the strings in $S$, at the same position as $s_{ci}$. The QUBO formulation is then presented incorporating two conflicting constraints, which are used to counter-act on the effect of the other. Ultimately, in order to eliminate the need for computing a piecewise function which involves an additional step while preparing the embedding of the problem when using the Kronecker-delta, another QUBO formulation is given, building on the fact that digital computers represent symbols in a numerical form. However, it is worthwhile to emphasize that this is not to gain a speedup in the optimization process, but to eliminate an additional step during the embedding.\par

    \subsection{Reformulation: Horizontal Minimzation of the Sum of Hamming Distances}
        Referring to the definition given for the \acrshort{acr_csp} by the Equations (\ref{eq:intro:hamming_distance}), (\ref{eq:intro:hamming_distance_f}) and (\ref{eq:intro:csp_condition}), it is established that $k$ in Equation (\ref{eq:intro:csp_condition}) should be minimized. Consider that, as given above, we have chosen a string $s_{c}$ over alphabet $\Sigma$, as a candidate solution to the \acrshort{acr_csp}. Let $D(s_{c})$ be the sum of Hamming distances of the string $s_{c}$ with the strings $s_{x} \in S$. Therefore, we can define $D(s_{c})$ as follows.
        \begin{equation} \label{eq:formulation:sum_ham_dist}
            D(s_{c}) = \sum_{x = 1}^{n} d(s_{c}, s_{x}) = d(s_{c}, s_{1}) + d(s_{c}, s_{2}) + ... + d(s_{c}, s_{n})
        \end{equation}
        According to Equation (\ref{eq:intro:csp_condition}), following extended constraint can be deduced. For $s_{M}$,
        \begin{equation} \label{eq:formulation:condition_extended_csp}
            d(s_{M}, s_{1}) + d(s_{M}, s_{2}) + ... + d(s_{M}, s_{n}) \le nk
        \end{equation}
        That is,
        \begin{equation} \label{eq:formulation:condition_extended_csp_as_obj}
            D(s_{M}) \le nk
        \end{equation}
        Hence, the \acrshort{acr_csp} can be restated as: Determine a string $s_{M}$ that minimizes $k$ in Equation (\ref{eq:formulation:condition_extended_csp_as_obj}). Here, in order to minimize $k$, it is required to minimize $D(s_{c})$ and,
        \begin{equation}
            argmin_{s_{c}}\left\{D(s_{c})\right\}_{s_{c} \in \Sigma^{m}} = s_{M}
        \end{equation}
        where $\Sigma^{*}$ is the language containing all the strings of length $m$ over the alphabet $\Sigma$.\par

        If the strings $s_{x} \in S$ are arranged in an $n \times m$ matrix, where each row correspond to a string, then the objective in the Equation (\ref{eq:formulation:condition_extended_csp_as_obj}) above can be portrayed as minimizing the Hamming distance between the candidate solution and the strings in the rows of the matrix - horizontally.\par

    \subsection{Reformulation: Vertical Minimization of the Sum of Hamming Distances}
        Let the string $s_{c}$ and $s_{x}$ be expressed as $s_{c} = s_{c1}s_{c2}...s_{cm}$ and $s_{x} = s_{x1}s_{x2}...s_{xm}$, where $s_{ci}$ and $s_{xi}$ are the symbols at the $i^{th}$ position of the strings $s_{c}$, $s_{x}$ respectively. By using the definition in the Equation (\ref{eq:intro:hamming_distance}), we can expand the RHS of the Equation (\ref{eq:formulation:sum_ham_dist}) as follows.
        \begin{align*}
            D(s_{c}) &= \{ f(s_{c1}, s_{11}) + f(s_{c2}, s_{12}) + ... + f(s_{cm}, s_{1m}) \} \\
                     &+ \{ f(s_{c1}, s_{21}) + f(s_{c2}, s_{22}) + ... + f(s_{cm}, s_{2m}) \} \\
                     &+ ... \\
                     &+ \{ f(s_{c1}, s_{n1}) + f(s_{c2}, s_{n2}) + ... + f(s_{cm}, s_{nm}) \}
        \end{align*}
        Rearranging,
        \begin{align*}
            D(s_{c}) &= \{ f(s_{c1}, s_{11}) + f(s_{c1}, s_{21}) + ... + f(s_{c1}, s_{n1}) \} \\
                     &+ \{ f(s_{c2}, s_{12}) + f(s_{c2}, s_{22}) + ... + f(s_{c2}, s_{n2}) \} \\
                     &+ ... \\
                     &+ \{ f(s_{cm}, s_{1m}) + f(s_{cm}, s_{2m}) + ... + f(s_{cm}, s_{nm}) \}
        \end{align*}
        \begin{equation} \label{eq:formulation:rearranged_extended_csp}
            \therefore
            D(s_{c}) = \sum_{i = 1}^{m}\sum_{x = 1}^{n} f(s_{ci}, s_{xi})
        \end{equation}
        Since it is established that $D(s_{c})$ is required to be minimized in Equation (\ref{eq:formulation:condition_extended_csp_as_obj}), we can conclude, by considering the Equation (\ref{eq:formulation:rearranged_extended_csp}), that $\Delta_{i}(s_{c})$ for each $i = 1,2,...,m$ must be minimized, where $\Delta_{i}(s_{c})$ is defined as,
        \begin{equation} \label{eq:formuation:def_Delta_i}
            \Delta_{i}(s_{c}) = \sum_{x = 1}^{n} f(s_{ci}, s_{xi})
        \end{equation}
        This is achievable because there are no terms that correlates the distance between each symbol position $i$, as suggested by the Equation (\ref{eq:formulation:rearranged_extended_csp}). An important implication of this result is that it is possible to decompose \acrshort{acr_csp} to $m$ sub-problems that can be solved independently.\par

        From the Equations (\ref{eq:formulation:rearranged_extended_csp}) and (\ref{eq:formuation:def_Delta_i}), it is evident that the \acrshort{acr_csp} can be restated as follows: For each symbol position $i = 1,2,...,m$, determine the symbol $s_{ci}$ which minimizes $\Delta_{i}(s_{c})$. i.e., for every $i$, $\Delta_{i}(s_{M})$ is minimum.\par

        Following the same matrix arrangement as in the previous section, this formulation can be portrayed as the iterative minimization of the Hamming distance between string generated by repeating the $i^{th}$ symbol of the candidate solution $n$ times and the string in the $i^{th}$ column of the matrix, for each $i$ - vertically.\par

    \subsection{Reduced Search Space for each Symbol Position}
        The candidate solution $s_{c}$ is defined as a string over the same alphabet $\Sigma$, which is the alphabet for the strings in set $S$. However, according to the definition given in the Equation (\ref{eq:intro:hamming_distance_f}), if $s_{ci} \notin \Sigma_{i}$ then, for all $x = 1,2,...,n$, $f(s_{ci}, s_{xi}) = 1$, where $\Sigma_{i} = \{s_{xi} : s_{xi} \text{ is the symbol at the }i^{th}\text{ position of the string }s_{x} \in S\}$ for each $i$. Therefore, the following can be stated. If $\omega \in \Sigma_{i}$, $\omega^{'} \notin \Sigma_{i}$ and $s_{c^{'}}$ is a string defined over $\Sigma$ where $s_{c^{'}i} \notin \Sigma_{i}$ then,
        \begin{equation} \label{eq:formulation:reducing_ss_ineq_dist}
            \sum_{x = 1}^{n}f(\omega, s_{xi}) < \sum_{x = 1}^{n}f(\omega^{'}, s_{xi}) \implies \Delta_{i}(s_{c}) < \Delta_{i}(s_{c^{'}})
        \end{equation}
        It is guaranteed by the Equation (\ref{eq:formulation:reducing_ss_ineq_dist}) that the symbol at $i^{th}$ position of the string $s_{M}$ is one of the symbols at the $i^{th}$ position of any of the strings $s_{x} \in S$. Accordingly, the most important result here is that we can reduce the search space of the $s_{ci}$ to be the set $\Sigma_{i}$, instead of $\Sigma$. Note that $c^{'}$ in $s_{c^{'}}$ is used to emphasize that the entire string may not be different from $s_{c}$, but just a symbol (the use of ' in the subscript).\par

        Therefore, for a given $s_{ci}$ if $\Delta_{i}(s_{c})$ for a given symbol position $i$ is minimum then $s_{Mi} = s_{ci}$ is probable. Assuming the minimality of $\Delta_{i}(s_{c})$, if $\nexists \omega \in \Sigma_{i}$ such that $s_{ci} \neq \omega$ and $\Delta_{i}(s_{c}) = \Delta_{i}(\omega)$, then $s_{Mi} = s_{ci}$ is guaranteed. Conclusively, we can state the following. For each $i = 1,2,...,m$,
        \begin{equation} \label{eq:formuation:final_argmin}
            argmin_{x}\left\{\sum_{j = 1}^{n}f(s_{xi}, s_{ji})\right\}_{x = 1}^{n} = c_{Mi}
        \end{equation}

    \subsection{QUBO Formulation of the \acrshort{acr_csp}}
        According to the result given by the Equation (\ref{eq:formulation:reducing_ss_ineq_dist}), it is guaranteed that the symbol corresponding to the minimum distance for each symbol position $i$ exists in $\Sigma_{i}$. Therefore, per each symbol position, allocating $n$ binary variables (to account for each string), the following Hamiltonian results.
        \begin{equation} \label{eq:formulation_qubo_1_H_B}
            H_{B} = B \sum_{i = 1}^{m} \sum_{x = 1}^{n} \alpha_{xi} \sum_{y = 1}^{n} f(s_{xi}, s_{yi})
        \end{equation}
        Here, the binary variable $\alpha_{xi}$ denotes whether the symbol $s_{xi}$ is chosen as the symbol $s_{ci}$ in the candidate solution $s_{c}$. $B$ is the Lagrange multiplier\footnote{This is identified as a Lagrange parameter in D-Wave Systems\cite{noauthor_getting_nodate}}. The Equation (\ref{eq:formulation_qubo_1_H_B}) represents the objective term.\par

        The ground state of the Hamiltonian $H_{B}$ alone corresponds to the state where for all $x = 1,2,...,n$ and $i = 1,2,...,m$, $\alpha_{xi} = 0$. Not all variables should take the value 0. Optimally and ideally, for every $x$, only one variable in the set $\{\alpha_{x1}, \alpha_{x2}, ..., \alpha_{xm}\}$ is allowed to take the value $1$, as only one symbol is to be chosen for $s_{Mi}$. Evidently, these two constraints conflict with each other. Nevertheless, by introducing penalty terms to the Hamiltonian in the Equation (\ref{eq:formulation_qubo_1_H_B}), the lowest energy can still be attributed to the the state which satisfies both of the constraints. Accordingly, following Hamiltonian represents both of the penalty terms.
        \begin{equation} \label{eq:formulation_qubo_1_H_A}
            H_{A} = A \sum_{i = 1}^{m} \sum_{x = 1}^{n} \left\{1 - \alpha_{xi}\right\} + A \sum_{i = 1}^{m} \sum_{x = 1}^{n} \left\{\alpha_{xi} \sum_{y = x + 1}^{n} \alpha_{yi}\right\}
        \end{equation}
        Here, the first penalty term accounts for the constraint that disallows all variables from taking the value 0. In fact, it adds up a penalty for every variable that is set to 0. Including quadratic terms progressively for each pair of variables, per each symbol position, the second penalty term increases the penalty assigned for configurations in which there are more than one variable with value 1, for each symbol position. $A$ is the Lagrange multiplier associated with the Hamiltonian $H_{A}$. Notably, the same Lagrange multiplier was used for both of the penalty terms, signifying that both of the constraints are of equal importance in determining $s_{M}$.\par

        By combining the Hamiltonians given in the Equations (\ref{eq:formulation_qubo_1_H_A}) and (\ref{eq:formulation_qubo_1_H_B}), it can be concluded that the following Hamiltonian $H$ describes the \acrshort{acr_qubo} formulation for \acrshort{acr_csp}.
        \begin{equation} \label{eq:formulation_qubo_1_H}
            H = H_{A} + H_{B}
        \end{equation}

    \subsection{An Alternative \acrshort{acr_qubo} Formulation}
        Even though the optimization strategy in quantum annealing is not constructed upon a gradient-based optimization algorithm, it is generally preferred to avoid conditional statements in the problem formulation of an optimization problem as they result in discontinuous functions\cite{parkinson_optimization_2013}. Additionally, despite the fact that problem embedding procedure is not performance critical and being done by using a modern programming language with a compiler that is optimized to handle logical operations as efficiently as the arithmetic operations (or vice-versa depending on the exact architecture being used), one might prefer to formulate the problem in the form of a continuous function. Based on the internal numerical representation of symbols in digital computers\footnote{By using encoding standards such as ASCII}, this is achievable. The formulated function will not be continuous by its definition, but will be entirely based on arithmetic operations. Let $C: \Sigma \rightarrow \mathbb{R}$ be a bijection and $C(s_{xi})$ denote the value of the mapping for the symbol at $i^{th}$ position of the string $s_{x}$. Using this mapping $C$, the following Hamiltonian $H_{B}^{'}$ can be used as an alternative objective function instead of the Hamiltonian given by the Equation (\ref{eq:formulation_qubo_1_H_B}).
        \begin{equation} \label{eq:formulation_qubo_2_H_B}
            H_{B}^{'} = B \sum_{i = 1}^{m}\sum_{x = 1}^{n} \alpha_{xi} \sum_{y = 1}^{n} \frac{\left(C(s_{xi}) - C(s_{yi})\right)^{2}}{\left(C(s_{xi}) - C(s_{yi})\right)^{2} + 1}
        \end{equation}
        Notice that in this case, each value is scaled to the range $[0,1)$. However, this is to facilitate tuning of the Lagrange multipliers whereas in most of the classical optimization algorithms it is to trigger faster convergence and to avoid biases\cite{parkinson_optimization_2013}.\par

        Similar to the Equation (\ref{eq:formulation_qubo_1_H}), combining Hamiltonians given in the Equations (\ref{eq:formulation_qubo_1_H_A}) and (\ref{eq:formulation_qubo_2_H_B}), we arrive at the following alternative formulation for \acrshort{acr_csp}, where $H^{'}$ gives the Hamiltonian.
        \begin{equation} \label{eq:formulation_qubo_2_H}
            H^{'} = H_{A} + H_{B}^{'}
        \end{equation}

    \subsection{Choosing Values for the Lagrange Parameters}
        In both of the Hamiltonians given by the Equations (\ref{eq:formulation_qubo_1_H}) and (\ref{eq:formulation_qubo_2_H}), it is required to determine values for the Lagrange multipliers $A$ and $B$. Evidently, they could be plugged in with arbitrary values based on the requirements of the problem of interest. It depends on whether or not it is feasible to violate constraints at the cost of further minimizing the objective. The following guidelines are followed in this regard.\par

        It is advisable to set $B = 1$. The constraints specified above must never be violated. Therefore, it must be assured that $max\{H_{B}\} < min\{H_{A}\}$. Since $max\{H_{B}\} = Bmn(n - 1)$, we need to ensure that $min\{H_{A}\} > mn(n - 1)$. In order to determine $min\{H_{A}\}$, its two terms can be considered independently by assigning either $\alpha_{xi} = 0$ or $\alpha_{xi} = 1\text{ }\forall x,i$, to see that
        \begin{equation} \label{eq:choose_lp_min_H_A}
            min\{H_{A}\} =
                \begin{cases}
                    Am, & \text{if } n = 2\\
                    \frac{Am(n - 1)(n - 2)}{2}, & \text{if } n = 3\text{ or }n = 4\\
                    Amn, & \text{if } n > 4
                \end{cases}
        \end{equation}
        The proof of the Equation (\ref{eq:choose_lp_min_H_A}) is given in the Appendix. Considering the distribution of the symbols in each position from $1$ to $m$, $\exists A \in \mathbb{R}$ such that 
        $$B < A \le \left\lceil \frac{Bmn(n - 1)}{\lambda} \right\rceil$$
        where, 
        $$
            \lambda = \frac{min\{H_{A}\}}{A} = 
                \begin{cases}
                    m, & \text{if } n = 2\\
                    \frac{m(n - 1)(n - 2)}{2}, & \text{if } n = 3\text{ or }n = 4\\
                    mn, & \text{if } n > 4
                \end{cases}
        $$ 
        which guarantees the optimal solution. Hence, for the purpose of analysis, a range of values from $\left(B, \left \lceil \frac{Bmn(n - 1)}{\lambda} \right \rceil \right]$ can be attempted for $A$. If, by inspection, it appears that all the strings are mostly similar, then it is recommended to start with a value closer to the lower bound of possible values for $A$. i.e., closer to $B$. Otherwise, start with a value closer to the upper bound of $A$.\par

        When using the Hamiltonian $H^{'}$, it is pertinent to observe that the values for these Lagrange parameters could be slightly different, typically with a minimal impact on the results. However, this minor difference assumes significance when the obtained solutions of $H$ are slightly different from the solutions of $H^{'}$.\par

    \subsection{Important Concerns in the D-Wave Systems}
        Each quadratic term in either of the \acrshort{acr_qubo} formulations given in Equation (\ref{eq:formulation_qubo_1_H}) and Equation (\ref{eq:formulation_qubo_2_H}) translates to a connection between a pair of qubits in the working graph. In the minor embedding, these two qubits corresponds to the two binary variables in the quadratic term concerned. For every $i = 1,2,...,m$, the interactions between all pairs of terms in the second constraint(term) of $H_{A}$ can be represented by an undirected complete graph $K_{n}$. According to \cite{boothby_next-generation_2020}, allowing longer chains, it is possible to derive minor-embeddings for up to $n = 12M - 10$, where $M$ is the working graph size in the Pegasus topology. Advantage4.1, which was the default, publicly available and the preferred \acrshort{acr_qpu} choice at the time of experimentation, has a graph size of 16 \cite{noauthor_qpu-specific_nodate}. i.e., $M = 16$. Therefore, the the upper bound is $n = 182$. If each one of the $m$ sub-problems is to be solved independently, Advantage4.1 \acrshort{acr_qpu} with P16 (Pegasus working graph with 16x16 unit cells) graph will be able to solve \acrshort{acr_csp} instances with a maximum number of strings of 182. For any other instance, if $m_{max}$ is the maximum number of sub-problems that is minor-embedded at once, then the maximum number of strings is $\lfloor182/m_{max}\rfloor$. Subsequently it can be concluded that if the number of strings in a given \acrshort{acr_csp} exceeds $182$, it cannot be solved with a \acrshort{acr_qpu} leveraging the P16 graph.\par

        As briefly explained in the preliminaries, given a problem graph $Q$ that is obtained by representing the variables in the \acrshort{acr_qubo} formulation as vertices and interactions between the variables given by the quadratic terms as edges, typically it is required to embed $Q$ in the working graph $F$ as a minor. However, there are instances where there exists a one-to-one mapping of vertices $f$ from graph $Q$ to graph $F$ where each edge of $Q$ is an edge of $F$. If such a mapping $f$ exists, then $Q$ is a subgraph of $F$ and a subgraph embedding is possible\cite{klymko_adiabatic_2014}. In such cases, the optimum embedding does not contain chains. $K_{4}$ is a subgraph of the working graph of the P16\cite{boothby_next-generation_2020}. Accordingly, all problem graphs $Q = K_{q}$ such that $1 \le q \le 4$ has a subgraph embedding in $P$, where $P$ denotes the working graph of P16. Consequently, all instances of \acrshort{acr_csp} such that $n \le 4$ has a subgraph embedding. Therefore, whenever the number of strings ($n$) is less than or equal to $4$ for the given problem instance, ``chain\textunderscore strength" parameter can be set to $0$. However, for other cases a non-zero value is required, which can be assigned by considering the maximum possible value for the \acrshort{acr_qubo}.\par

\section{Experimental Evaluation using D-Wave Systems}
    D-Wave systems’ Leap cloud has been utilized as the platform and the Python \acrshort{acr_sdk} was used to build a script that would embed and solve the problem\footnote{Codebase: https://github.com/chandeepadissanayake/qubo\textunderscore csp}. Refer to the preliminaries for further explanation on the D-Wave annealers. It is worthwhile to emphasize that the purpose of this evaluation is simply to establish empirical evidence for the validity of \acrshort{acr_qubo} formulations. This would not serve as a rigorous or exhaustive testing attempt across many possible test cases.\par

    Both QUBOs were tested for 4 sets of strings. The first 3 sets were handcrafted and the last one was created by ChatGPT\cite{noauthor_introducing_nodate}. Refer to the Table~\ref{table:eval_strings} for the sets of strings. Each set of strings was evaluated separately. For each set, ``num\textunderscore reads" was set to 100. Furthermore, when implementing Hamiltonian $H^{'}$ in the Equation (\ref{eq:prelim:qubo_2}), ASCII representation was used as the bijection $C$.\par
    \begin{table}
        \centering
        \caption{The expected closest string for each set of strings}
        \label{table:eval_strings}
        \begin{tabular}{|c|p{5cm}|p{2cm}|}
            \hline
            \textbf{Set} & \textbf{Set of Strings} & \textbf{Expected Closest String} \\
            \hline
            \#1 & \{``aaa",``aaa",``ddd"\} & ``aaa" \\
            \hline
            \#2 & \{``aaa",``aaa",``ddd",``ddd",``ddd"\} & ``ddd" \\
            \hline
            \#3 & \{``aaa",``aaa",``ded",``ded",``ded",``ddd"\} & ``ded" \\
            \hline
            \#4 & \{``abcdef",``ghijkl",``abcghi",``xyzjkl", ``abcmno"\} & ``abcjkl" \\
            \hline
        \end{tabular}
    \end{table}
    \par
    
    \subsection{Determining the Chain Strength}
        Since Set \#1 has only 3 strings, it allowed for the ``chain\textunderscore strength" parameter value to be set to 0. However for the other sets, it has to be assigned to a non-zero value as the number of strings in the set exceeded 4. The exact value assigned was dependent on the number of strings in the respective set and the distribution of the symbols in each set. By considering the influence of these two factors, we can discern four distinct cases which would determine the value for the ``chain\textunderscore strength", which is enumerated in ascending order as follows.
        \begin{enumerate}
            \item Less number of strings with a low variance in the distribution of symbols in the strings \label{exp_dwave_cs_case_1}
            \item Less number of strings with a high variance in the distribution of symbols in the strings \label{exp_dwave_cs_case_2}
            \item Higher number of strings with a low variance in the distribution of symbols in the strings \label{exp_dwave_cs_case_3}
            \item Higher number of strings with a high variance in the distribution of symbols in the strings \label{exp_dwave_cs_case_4}
        \end{enumerate}
        It can be observed that out of the two factors, the number of strings takes precedence over the variance in the distribution of symbols within the strings. Increasing the number of strings directly leads to an increased number of chains in the minor-embedding. While it may not be immediately evident that a high variance in symbol distribution directly causes chain breaks, it indirectly leads to a greater contribution to the overall energy within the Hamiltonian, thus diminishing the significance of the terms representing the chains. The comparison w.r.t. the number of strings and the distribution of symbols implied within the cases outlined above are relative to a given baseline. One such evident baseline, which is used in this work, is $\text{``chain\textunderscore strength"} = 0$ for $n \le 4$ as elaborated above. Each and every case above refers to $n > 4$. For the case \ref{exp_dwave_cs_case_1}, the general heuristic is to start off with a slightly higher value than the baseline. Accordingly, a preferred starting point for case \ref{exp_dwave_cs_case_1} is $\text{``chain\textunderscore strength"} = 1$. For the subsequent cases \ref{exp_dwave_cs_case_2}, \ref{exp_dwave_cs_case_3}, \ref{exp_dwave_cs_case_4}; values with increasing magnitude should be used. Accordingly, for the string set \#1, baseline value is used and in the sets \#2, \#3 and \#4, which can be considered as problem instances pertaining to the cases \ref{exp_dwave_cs_case_2}, \ref{exp_dwave_cs_case_3} and \ref{exp_dwave_cs_case_4} respectively, a range of values for ``chain\textunderscore strength" is attempted and the value that produced the optimum results is reported. It is worth noting that, despite the D-Wave documentation specifying that the ``chain\textunderscore strength" value is considered a hyperparameter\cite{noauthor_programming_nodate}, it is beneficial to establish a set of guidelines to arbitrarily limit the range of possible values.\par

    \subsection{Occurrence Ratio of a Solution}
        Due to the intrinsic probabilistic nature of the \acrshort{acr_qpu}s in D-Wave systems, the optimal solution may not be returned in every annealing cycle. Based on the \acrshort{acr_qubo} formulations in the Equations (\ref{eq:formulation_qubo_1_H}) and (\ref{eq:formulation_qubo_2_H}), it is evident that multiple configurations of output can present the same solution in many cases. Therefore, in order to determine the significance of a solution among the other solutions in the solution space, a metric is desired. Note that the different configurations of output which present the same output string as the solution must be treated as the respective output string occurring multiple times. For a given solution/output string P, the Occurrence Ratio $OR_{P}$ is defined as follows.
        \begin{equation} \label{eq:exp_dwave_or_def}
            OR_{P} = \frac{N_{P}}{\sum_{Q}N_{Q}}
        \end{equation}
        where, $N_{P}$ represents the number of occurrences of $P$ in the solution space. It should be evident that $\sum_{Q}N_{Q} = \text{num\textunderscore reads}$.\par
        
        A higher occurrence ratio implies a higher likelihood of the solution being obtained from a D-Wave annealer. Maximum Occurrence Ratio ($MOR$) is the maximum of all possible $OR_{P}$ values, i.e., the maximum $OR_{P}$ that happens for any solution. $MOR$ will aid in comparing the likelihood of the most occurring solution in the solution space against the likelihood of a given solution $P$.\par

    \subsection{Results}
        \begin{table}[!t]
            \renewcommand{\arraystretch}{1.3}
            \centering
            \caption{Results using Hamiltonian $H$ on D-Wave Leap Cloud QPUs}
            \label{tab:exp_dwave_results_H}
            \begin{tabular}{|c|c|c|c|c|c|c|}
                \hline
                Set \# & $P$ & $A$ & $B$ & $\gamma$ & $OR_{P}$ & $MOR$ \\
                \hline
                \#1 & “aaa” & 2 & 1 & 0 & 1.00 & 1.00 \\
                \#2 & “ddd” & 3 & 1 & 1 & 0.99 & 0.99 \\
                \#3 & “ded” & 5 & 1 & 6 & 0.53 & 0.53 \\
                \#4 & “abcjkl” & 4 & 1 & 5 & 0.22 & 0.22 \\
                \hline
            \end{tabular}
        \end{table}

        \begin{table}[!t]
            \renewcommand{\arraystretch}{1.3}
            \centering
            \caption{Results using Hamiltonian $H^{'}$ on D-Wave Leap Cloud QPUs}
            \label{tab:exp_dwave_results_H_}
            \begin{tabular}{|c|c|c|c|c|c|c|}
                \hline
                Set \# & $P$ & $A$ & $B$ & $\gamma$ & $OR_{P}$ & $MOR$ \\
                \hline
                \#1 & “aaa” & 2 & 1 & 0 & 1.00 & 1.00 \\
                \#2 & “ddd” & 3 & 1 & 1 & 0.97 & 0.97 \\
                \#3 & “ded” & 5 & 1 & 6 & 0.51 & 0.51 \\
                \#4 & “abcjkl” & 4 & 1 & 5 & 0.22 & 0.22 \\
                \hline
            \end{tabular}
        \end{table}

    Refer to the Table~\ref{tab:exp_dwave_results_H} and Table~\ref{tab:exp_dwave_results_H_} for the results when the \acrshort{acr_qubo} formulations  given by the Equations (\ref{eq:formulation_qubo_1_H}) and (\ref{eq:formulation_qubo_2_H}) are used. Column ``$P$" refers to the minimum energy solution obtained for each set of strings. The  columns ``$A$", and ``$B$" refer to the corresponding values for the respective Lagrange parameters whereas column ``$\gamma$" refers to the value for ``chain\textunderscore strength" used. ``$OR_{P}$" and ``$MOR$" columns hold the same definitions as defined in the preceding section. The minimum energy solution $P$ has always been reported disallowing any chain breaks by setting the appropriate value for ``chain\textunderscore strength" $\gamma$ while repeating ``num\textunderscore reads" ($ = 100$), over several attempts to determine the maximum $OR_{P}$ value possible for $P$.\par

    Since sets \#1 and \#2 contained similar strings within them, values closer to the lower bound of $A$ were used. Even though the set \#3 contained similar strings, a larger number of strings suggested the possibility of a constraint violation if a lesser value is used for $A$, which was confirmed by running it with $A = 2$. The outcome of this case is not reported as it was not optimal. Instead, with slightly a higher value, i.e., by setting $A = 5$ the expected solution was obtained. Set \#4, contained similar, yet slightly different strings with a comparatively concentrated distribution of symbols, prompting the test case to be done with slightly a lesser value for $A$. Conclusively, in all cases, the expected closest string has been obtained and the MOR is minimal for every solution. It is worthwhile to note that given the random nature of the QPUs, these values for OR/MOR may not be exactly reproducible, but proximal outcomes should be expected.\par

    It can be observed that the $OR_{P}$ values for sets \#3 and \#4 are comparatively lower than what was obtained for sets \#1 and \#2. This can mostly be attributed to the largely narrow energy landscape created by the narrow distribution of the symbols in the sets \#1 and \#2. The requirement of a higher ``chain\textunderscore strength" in the cases \#3 and \#4 has augmented this effect by dispersing out a narrow distribution, possibly into a one with many local minima. It is possible that tuning of the Lagrange multipliers and the ``chain\textunderscore strength" could have alleviated this effect to some extent, but it was not a concern in this work. It should also be emphasized that for a noisy optimization process, it is cumbersome to point out the exact cause.\par

\section{Discussion}
    One of the most interesting aspects of both of the \acrshort{acr_qubo} formulations is that they have resulted in formulations where there is no interaction between qubits representing symbols at different positions across the strings. Therefore, the problem can be decomposed to sub-problems at the level of individual symbols. Thus, such sub-problems can be independently solved on the \acrshort{acr_qpu}. For a larger number of strings, this approach could be followed, decomposing the strings to substrings of equal size, which can be directly embedded in the \acrshort{acr_qpu} at once.\par

    Consequently, the number of symbols in the strings does not affect the possibility of being solvable on a given \acrshort{acr_qpu} architecture. However, the number of strings remains to be a limiting factor. In fact, as calculated previously, an instance of \acrshort{acr_csp} with a maximum of 182 strings can be solved with P16 \acrshort{acr_qpu} architecture in D-Wave systems. Theoretically, the number of qubits required under both \acrshort{acr_qubo} formulations is $mn$. However, in the P16 architecture minor-embedding can lead up to chains with the length 16/17 for a single node in the problem graph and hence the exact number of qubits required is strictly dependent on the architecture of the \acrshort{acr_qpu}.\par

\section{Conclusions and Recommendations}
    Both of the derived \acrshort{acr_qubo} formulations can be used for solving the \acrshort{acr_csp} on a quantum annealer. For \acrshort{acr_csp}, the given constraints should not be violated and hence during hyperparameter tuning, the constraints should be strictly enforced. Guidelines for tuning the Lagrange parameters and the ``chain\textunderscore strength" have been provided and one may significantly improve the results by following these guidelines. The constraints specified above could be relaxed for different variants of the problem. Based on algorithms such as the greedy heuristic algorithm in \cite{vilca_recursive_2022}, it might be possible to investigate different \acrshort{acr_qubo} formulations for the \acrshort{acr_csp}, possibly relying on the global evaluation of the Hamming distance. An interesting experimental direction for future work based on these formulations is to evaluate the behavior of the minimum energy and the corresponding solutions across the domain of the values permitted for the Lagrange parameters, which may provide insights on hyperparameter tuning for sets of strings in different scales.\par

\section{Acknowledgments}
    The author would like to express his heartfelt gratitude to Dr. Chinthanie Weerakoon, a Senior Lecturer at the Department of Statistics and Computer Science, University of Kelaniya, for her unwavering support and guidance in the publication of this paper. Furthermore, he wants to extend his thanks to Dr. Anuradha Mahasinghe, a Senior Lecturer at the Department of Mathematics, University of Colombo, for mentoring him and sparking his interest in the adiabatic model of quantum computation. He also wishes to thank Dr. Sachintha Pitigala, a Senior Lecturer at the Department of Statistics and Computer Science, University of Kelaniya, for his support to the author in the academic arena.\par

\printbibliography

\begin{IEEEbiography}[{\includegraphics[width=1in,height=1.25in,clip]{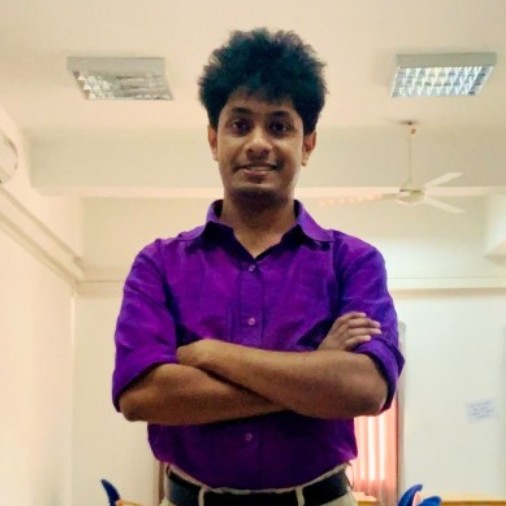}}]{Chandeepa Dissanayake}
is a senior undergraduate student of computer science at the University of Kelaniya, Sri Lanka. He has research interests in theoretical computer science, quantum computing, and quantum information science. He was awarded the second prize at the IEEE SA P2834 Student Challenge 2022 and several national awards for his innovative projects in computer science.
\end{IEEEbiography}

\end{document}